\documentclass[showpacs,amssymb]{revtex4}
\usepackage{mathrsfs}

\newcommand{\inv}{^{-1}}
\newcommand{\Tr}{\mathrm{Tr}}

\newcommand{\U}{\mathrm{U}}
\newcommand{\SU}{\mathrm{SU}}

\newcommand{\vect}[1]{\mbox{\boldmath $#1$}}
\newcommand{\diag}{\mathrm{diag}}

\begin{document}

\title{A Possible Generalization of Quantum Mechanics}
\author{Yu Tian}\email{ytian@bit.edu.cn}
\affiliation{Department of Physics, Beijing Institute of Technology,\\
Beijing 100081, P. R. China}
\date{\today}

\begin{abstract}
A ``minimal" generalization of Quantum Mechanics is proposed,
where the Lagrangian or the action functional is a mapping from
the (classical) states of a system to the Lie algebra of a general
compact Lie group, and the wave function takes values in the
corresponding group algebra. This formalism admits a probability
interpretation and a suitable dynamics, but has no obvious
classical correspondence. Allowing the Lagrangian or the action
functional to take values in a general Lie algebra instead of only
the real number field (actually the $\mathfrak{u}(1)$ algebra)
enlarges the extent of possible physical laws that can describe
the real world. The generalized quantum dynamics of a point
particle in a background gauge field is given as an example, which
realizes the gauge invariance by a Wilson line structure and shows
that some Schr\"odinger-like equation can be deduced within this
formalism. Some possible developments of this formalism are also
discussed.
\end{abstract}


\maketitle

Quantum Mechanics (QM) has become one of the most important
foundations of physics ever since its establishment in the 1920s.
While QM was achieving great success, physicists had begun to
consider the possibility to generalize the traditional framework
of it \cite{Jordan}. Up to now, although various generalizations
of QM are all not very useful and/or successful, this kind of
attempts have never stopped (see, for example, Refs.
\cite{Finkelstein1,Finkelstein2,LAK,MT} and references therein).
This letter proposes a new way to generalize QM, which makes few
changes in the formalism of QM and may be viewed as a minimal
generalization of QM. The greatest change caused by this
generalization may be that it renders QM to have no obvious
classical correspondence. However, this may not be a shortcoming
of the generalized QM, but rather an advantage in some sense. In
fact, although in the present framework of quantum theory
(including also quantum field theory and even string theory) it is
taken for granted that there should first be a classical system
and then it is quantized to give real physics, it has long been
suspected that our universe may eventually be governed by some
intrinsic quantum structure, i.e., there may be no corresponding
classical structure for the ultimate law of nature at all. The
generalized QM here may play such a role. As a simple example, the
generalized quantum dynamics of a point particle in a background
gauge field is proposed, which has some advantage over the
ordinary treatments. This example shows that the generalized
fundamental principles of QM can lead to some Schr\"odinger-like
equation with a formal Hamiltonian, but it is not the ordinary
quantization of a Hamilton's mechanics because the generalized
quantum system has no real-valued Lagrangian. Interestingly, the
gauge invariance in this model is naturally realized by a Wilson
line structure of the interaction.

In order to see why and how we can generalize QM, first let us
briefly review the fundamental principles of (non-relativistic)
QM. There are many versions of these principles. Here we only list
one from the viewpoint of the Schr\"odinger
picture:\footnote{Neither the strictness in mathematics nor the
axiomatization of physics is pursued here. These principles are
listed only for the purpose of generalization.}
\begin{enumerate}
    \item The quantum state of a system is described by the wave function
    $\psi$,\footnote{Here we do not use the notion ``ray".
    In fact, the usual definition of ray should be modified under our generalization.}
    which is a complex function of time $t$ and the canonical
    coordinates (collectively denoted by $\vect{r}$) of the
    system; at time $t$, the probability to find the system in the classical
    configuration $\vect{r}$ is proportional to $|\psi(t,\vect{r})|^2
    d\vect{r}$. \label{r probability}
    \item A mechanical quantity (observable) $F$ is a Hermitian operator acting on the Hilbert
    space formed by the set of wave functions, which has
    a complete set of eigenfunctions; the only possible result of one measurement of
    $F$ is one of its eigenvalues $\{f_n\}$.
    \item If the wave function of the system is the $n$-th
    eigenfunction $\psi_n$ of $F$, then the result of any
    measurement of $F$ is the corresponding eigenvalue $f_n$; if
    the wave function is a linear superposition $\sum_m
    a_m\psi_m$, then the probability to get a measuring result
    $f_n$ is $|a_n|^2/\sum_m|a_m|^2$. \label{F probability}
    \item The dynamics of the wave function is governed by the Schr\"odinger equation,
    which is a linear differential equation. \label{Schrodinger eq}
    \item The hypothesis of wave-packet collapse after
    measurements, whose details are not relevant here, or other
    measurement theories for QM.
\end{enumerate}
These principles are easy to understand and sufficient to
establish the whole framework of QM. A useful alternative of
Principle \ref{Schrodinger eq} is the so-called Feynman's
hypothesis of path integral. It is well known that there are
difficulties to set up the Feynman path integral on firm
mathematical foundation. However, a version of the Feynman path
integral in terms of infinitesimal time evolution seems less
troublesome. The dynamics of the wave function is described as a
Huygens-like principle, which determines the wave function at time
$t+\epsilon$ from the wave function at time $t$ as
\begin{equation}\label{Huygens}
\psi(t+\epsilon,\vect{r})=\int d\vect{r}'
K(t+\epsilon,\vect{r};t,\vect{r}')\psi(t,\vect{r}')
\end{equation}
with $K(t+\epsilon,\vect{r};t,\vect{r}')$ the propagator. Unlike
the Feynman's general hypothesis that assumes the propagator to be
a complicated path integral, for infinitesimal $\epsilon$ the
propagator is assumed to be ($\hbar=1$)
\begin{equation}\label{propagator}
K(t+\epsilon,\vect{r};t,\vect{r}')=N\exp\left[\mathrm{i}\epsilon
L\left(\frac{\vect{r}+\vect{r}'}{2},\frac{\vect{r}-\vect{r}'}{\epsilon},t\right)\right]
\end{equation}
with $L(\vect{r},\dot{\vect{r}},t)$ the Lagrangian and $N$ an
appropriate normalization factor. This infinitesimal version of
Feynman path integral is actually equivalent to the so-called
``polygonal paths" approach proposed by Feynman, and can be shown
to be equivalent to the Schr\"odinger equation.

As shown by the above fundamental principles, there are two kinds
of quantities that play important roles in QM. One is the wave
function, called also the probability amplitude, which is viewed
as the most fundamental; the other is the probability, which is
less fundamental but directly related to observations
(measurements). The probability interpretation (Principle \ref{r
probability} and \ref{F probability}) of QM just states that there
is a mapping from the former to the latter. Since the probability
must be a real number, there is no obvious way to generalize the
notion of probability. However, the notion of probability
amplitude is much more flexible, provided there exists a mapping
from it to the probability. That is the key point to make our
minimal generalization which keeps the other parts of QM almost
intact. In fact, many previous attempts to generalize QM just
follow this path \cite{Finkelstein1,Finkelstein2}.

Now the observation is that we can view the (usual) probability
amplitude as taking values in the $\U(1)$ group algebra, instead
of the complex number field as usual. These two viewpoints are
equivalent so far, but the one of group algebra accepts much
further generalization. Obviously, the next step is to allow the
probability amplitude to take values in a general Lie group
algebra, where the group is restricted to be compact for
simplicity in this letter. Moreover, we must define the mapping
from this group algebra to the probability so that the generalized
theory can adopt a probability interpretation. This can be easily
done with the aid of group representation theory. Given that any
compact Lie group $G$ has finite dimensional unitary
representations, and assuming the mapping to be quadratic, we can
define the mapping from any $g\in\mathcal{G}$ (the corresponding
group algebra of $G$) to the probability $p$ as
\begin{equation}
p(g)=\Tr(g^\dag g).
\end{equation}
Here (and in the following) we have used the same notation for $g$
and its representation, and the trace is taken in some fixed
representation. Compatible with the above mapping, an inner
product on the group algebra space can be immediately defined:
\begin{equation}
(g_1,g_2)=\Tr(g_1^\dag g_2),
\end{equation}
so that $p(g)$ can be identified with $(g,g)$. Alternatively, we
can use real representations (by orthonormal matrices) of $G$ and
define
\begin{equation}
(g_1,g_2)=\Tr(g_1^\mathrm{T} g_2),
\end{equation}
where the superscript T stands for matrix transposition.

In order to establish a complete set of fundamental principles as
before, we further assume that a mechanical quantity $F$ is still
a Hermitian operator on the ordinary Hilbert space, but not an
operator on the group algebra space. The time operator $t$ and the
coordinate operator $\vect{r}$, which have continuous spectrums of
eigenvalues, are also included in this statement. The whole wave
function space is thus a direct product of the ordinary Hilbert
space and the group algebra space. The original inner product on
the Hilbert space still works, which in general produces from the
whole wave function space a result on the group algebra space.

So far, our generalization is only notional. But it is indeed
possible for this generalization to have corresponding dynamics.
If we would like Eq. (\ref{Huygens}) to govern as usual the
dynamics of the generalized QM, where now the wave function
$\psi(t,\vect{r})$ takes values in $\mathcal{G}$, it is obvious
that the propagator $K$ can also take values in $\mathcal{G}$. If,
further, the infinitesimal propagator is still assumed to be given
by Eq. (\ref{propagator}), the Lagrangian should take values in
the Lie algebra $\mathfrak{g}$ of $G$. In other words, the
generalized Lagrangian is a mapping
\begin{equation}
L: (\vect{r},\dot{\vect{r}},t)\rightarrow\mathfrak{g}.
\end{equation}
Correspondingly, in the Feynman's general hypothesis of path
integral, the action functional should take values in
$\mathfrak{g}$. In fact, infinitely many successive propagators
(\ref{propagator}) can be sewn together to express the finite time
propagator as a generalized path integral
\begin{equation}\label{path integral}
K(t_2,\vect{r}_2;t_1,\vect{r}_1)=\int_{\vect{r}_1}^{\vect{r}_2}\mathcal{T}\left[\exp\int_{t_1}^{t_2}
L(\vect{r},\dot{\vect{r}},t) dt\right]\mathscr{D}\vect{r}(t)
=\int_{\vect{r}_1}^{\vect{r}_2}\mathcal{T}(e^{S[\vect{r}]})\mathscr{D}\vect{r}(t),
\end{equation}
where $\mathcal{T}$ denotes time ordering and the normalization
factor has been absorbed into the definition of
$\mathscr{D}\vect{r}(t)$. The above equation also shows that the
naive generalization of the Feynman path integral is incorrect
unless modified by the time ordering procedure. The Lagrangian
(action functional) in the ordinary QM takes values in
$\mathbb{R}$, which is just the Lie algebra of the $\U(1)$ group.
In this case the time ordering operator acts trivially, so the
generalized Feynman path integral becomes the ordinary one.

To sum up, our fundamental principles of the generalized QM, in
comparison with that of the ordinary QM stated above, are
\begin{enumerate}
    \item The same as before, but now $|\psi(t,\vect{r})|^2
    d\vect{r}$ is replaced with $p[\psi(t,\vect{r})]d\vect{r}$.
    \item The same as before.
    \item The same as before, but now $a_n$ takes values in the
    group algebra and $|a_n|^2$ is replaced with $p(a_n)$.
    \item The dynamics of the wave function is governed by
    Eqs. (\ref{Huygens}) and (\ref{propagator}), where $\psi$, $K$ and
    $L$ all have the generalized meaning.
    \item The same as before.
\end{enumerate}

A nice feature of our generalization is that a generalized quantum
system can be consistently coupled to an ordinary one simply
because the latter is a special case of the former. In fact, this
can be viewed as an important requirement for the generalization
to be physically acceptable, since most of the known physical
phenomena can be perfectly described by the ordinary quantum
theory. With the above feature, our generalized quantum theory may
describe some unclear physics or act as tiny corrections to known
physics. An interesting case is the $\U(N)$-generalized Lagrangian
\begin{equation}\label{U}
L=\mathrm{i} L_0+L_{\SU(N)},
\end{equation}
where $L_0$ is the ordinary real-valued Lagrangian and the
$\SU(N)$-valued part $L_{\SU(N)}$ acts as some perturbation.
Anyway, extending the range of Lagrangian or action functional to
a general Lie algebra enables one to have more choices of possible
physical laws for describing the real world. One may wonder how we
can have a natural mapping from a physical system to a general Lie
algebra. The answer is that physical systems themselves have Lie
algebras as symmetry structures.

Let us consider a point particle in a background gauge field. This
system was first considered by Wong \cite{Wong}, and then
discussed further by many other authors (see Ref. \cite{BSSW} and
references therein and thereafter). The generalized Lagrangian can
be taken as
\begin{equation}\label{Lagrangian}
L=\frac{\mathrm{i}}{2}(v\inv\dot{x}^2-v m^2)-A_\mu\dot{x}^\mu,
\quad \dot{x}^2=\eta_{\mu\nu}\dot{x}^\mu\dot{x}^\nu
\end{equation}
for relativistic case, where $\eta_{\mu\nu}=\diag(-1,1,1,1)$, $v$
is the Lagrange multiplier enforcing the mass-shell condition and
$A_\mu$ the anti-Hermitian gauge potential. The gauge group is
arbitrary, though it can be assumed as $\SU(N)$ in order for
comparison with Eq. (\ref{U}). Since the interaction term in Eq.
(\ref{Lagrangian}) commutes with the free part of the Lagrangian,
substitution of Eq. (\ref{Lagrangian}) into Eq. (\ref{path
integral}) shows that the interaction term just produces a Wilson
line factor in the path integral:
\begin{equation}
\int_{x_1}^{x_2}\mathcal{T}\left(\exp\int_{t_1}^{t_2} L
dt\right)\mathscr{D}x(t)=\int_{x_1}^{x_2} e^{\mathrm{i} S_0}
W[x;A]\mathscr{D}x(t), \quad W[x;A]=\exp\left(-\int_{t_1}^{t_2}
A_\mu dx^\mu\right),
\end{equation}
where $S_0$ is the (real-valued) free action functional. Under
gauge transformation
\begin{equation}
\tilde{A}_\mu=U A_\mu U\inv+U\partial_\mu U\inv,
\end{equation}
the Wilson line transforms as (see, for example, Ref.
\cite{Hatfield})
\begin{equation}
W[x;\tilde{A}]=U W[x;A] U\inv.
\end{equation}
Thus the propagator transforms similarly, which leads to the gauge
invariance of any physical results of this model. This realization
of the gauge invariance seems much easier than those before, where
additional degrees of freedom are inevitably introduced
\cite{BSSW}.

Further, to make our generalization physically interesting and to
show that the generalized Principle \ref{Schrodinger eq} can
really lead to some Schr\"odinger-like equation as before, we use
the above model as an example. For simplicity in the following
computation, we take the non-relativistic limit of the Lagrangian
(\ref{Lagrangian}) and restrict the spatial dimension to be one:
\begin{equation}\label{non-relativistic}
L(x,\dot{x},t)=\frac{\mathrm{i}}{2}m\dot{x}^2-A\dot{x}-\varphi,
\end{equation}
where we have defined $\varphi=A_0$ and $A=A_1$. Now we can apply
a standard approach (see, for example, Ref. \cite{FH}) to convert
Eq. (\ref{Huygens}) into the form of differential equation.
According to Eq. (\ref{propagator}) we have
\begin{eqnarray}
K(t+\epsilon,x;t,x')&=&N\exp\left[\mathrm{i}
m\frac{(x-x')^2}{2\epsilon}-(x-x') A\left(\frac{x+x'}{2}\right)
-\epsilon\varphi\left(\frac{x+x'}{2}\right)\right]\label{step1}\\
&\approx&N\exp\left[\frac{\mathrm{i}
m}{2\epsilon}\left(x'-x-\frac{\mathrm{i}\epsilon}{m}
A\right)^2\right] \left(1+\frac{\mathrm{i}\epsilon}{2m}
A^2-\epsilon\varphi\right).\label{step2}
\end{eqnarray}
From Eq. (\ref{step1}) to Eq. (\ref{step2}), the
Baker-Campbell-Hausdorff formula should be applied, but the extra
terms are all infinitesimals of higher order and so can be
neglected. Substitution of Eq. (\ref{step2}) into Eq.
(\ref{Huygens}) gives
\begin{equation}\label{Taylor}
\psi(t,x)+\epsilon\partial_t\psi(t,x)=N\int
d\xi\exp\left[\frac{\mathrm{i}
m}{2\epsilon}\left(\xi-\frac{\mathrm{i}\epsilon}{m}
A\right)^2\right] \left(1+\frac{\mathrm{i}\epsilon}{2m}
A^2-\epsilon\varphi\right)
\left[\psi(t,x)+\xi\partial_x\psi(t,x)+\frac{\xi^2}{2}\partial_x^2\psi(t,x)\right],
\end{equation}
where we have defined
\begin{equation}
\xi=x'-x
\end{equation}
and neglected all higher order terms in the Taylor expansions. The
zeroth order of Eq. (\ref{Taylor}) reads
\begin{equation}
N\int\exp\left(\frac{\mathrm{i} m}{2\epsilon}\xi^2\right) d\xi=1,
\end{equation}
so it follows that
\begin{eqnarray}
N\int\exp\left\{\frac{\mathrm{i}
m}{2\epsilon}\left[\xi-\frac{\mathrm{i}\epsilon}{m}
A\left(x+\frac{\xi}{2}\right)\right]^2\right\} d\xi &\approx&
N\int\exp\left\{\frac{\mathrm{i}
m}{2\epsilon}\left[\left(1-\frac{\mathrm{i}\epsilon}{2m}\partial_x
A\right)\xi-\frac{\mathrm{i}\epsilon}{m} A\right]^2\right\} d\xi
\\
&\approx& 1+\frac{\mathrm{i}\epsilon}{2m}\partial_x A.
\end{eqnarray}
Knowing also
\begin{equation}
N\int\xi\exp\left[\frac{\mathrm{i}
m}{2\epsilon}\left(\xi-\frac{\mathrm{i}\epsilon}{m}
A\right)^2\right] d\xi=N\int\frac{\mathrm{i}\epsilon}{m}
A\exp\left[\frac{\mathrm{i}
m}{2\epsilon}\left(\xi-\frac{\mathrm{i}\epsilon}{m}
A\right)^2\right] d\xi\approx\frac{\mathrm{i}\epsilon}{m} A
\end{equation}
and
\begin{eqnarray}
N\int\xi^2\exp\left[\frac{\mathrm{i}
m}{2\epsilon}\left(\xi-\frac{\mathrm{i}\epsilon}{m}
A\right)^2\right] d\xi &=&
N\int\left[\left(\xi-\frac{\mathrm{i}\epsilon}{m}
A\right)^2-\frac{\epsilon^2}{m^2}
A^2\right]\exp\left[\frac{\mathrm{i}
m}{2\epsilon}\left(\xi-\frac{\mathrm{i}\epsilon}{m}
A\right)^2\right] d\xi \\
&\approx&
N\int\frac{\mathrm{i}\epsilon}{m}\exp\left[\frac{\mathrm{i}
m}{2\epsilon}\left(\xi-\frac{\mathrm{i}\epsilon}{m}
A\right)^2\right] d\xi\approx\frac{\mathrm{i}\epsilon}{m}
\end{eqnarray}
from integration by parts, Eq. (\ref{Taylor}) then becomes
\begin{equation}\label{Schrodinger-like}
\hbar\partial_t\psi(t,x)=\left[\frac{\mathrm{i}\hbar^2}{2m}\partial_x^2
+\frac{\mathrm{i}\hbar}{m} A\partial_x+\frac{\mathrm{i}\hbar}{2m}
(\partial_x A)+\frac{\mathrm{i}}{2m} A^2-\varphi\right]\psi(t,x)
=\left[\frac{\mathrm{i}}{2m}(\hbar\partial_x+A)^2-\varphi\right]\psi(t,x),
\end{equation}
where we have restored the Planck constant omitted in Eq.
(\ref{propagator}). This Schr\"odinger-like equation is gauge
invariant if the wave function transforms as
\begin{equation}
\tilde{\psi}(t,x)=U\psi(t,x).
\end{equation}
Thus we have easily obtained a satisfactory quantum dynamics of
this system, while the ordinary quantization procedure is much
more complicated \cite{BSSW}.

Nevertheless, the Schr\"odinger-like equation
(\ref{Schrodinger-like}) can be viewed as a standard Schr\"odinger
equation with a multi-component $\psi(t,x)$ and a Hamiltonian
\begin{equation}\label{Hamiltonian}
\mathcal{H}=-\frac{1}{2m}(\hbar\partial_x+A)^2-\mathrm{i}\varphi.
\end{equation}
But the point is that the generalized quantum system does not have
a real-valued Lagrangian, so its classical correspondence is not
clear\footnote{In fact, it is easy to see that the ordinary
Euler-Lagrange equation deduced from the Lagrangian
(\ref{Lagrangian}) or (\ref{non-relativistic}) is inconsistent
unless the gauge potential takes values in $\mathfrak{u}(1)$.} and
the above Hamiltonian is just formal if it has nothing to do with
the (generalized) Lagrangian. In this example, we may formally
define the canonical momentum conjugate to $x$ as
\begin{equation}
p=\frac{\partial L}{\partial\dot{x}}=\mathrm{i} m\dot{x}-A,
\end{equation}
which takes values in the Lie algebra. Thus the ``classical
Hamiltonian" is
\begin{equation}
H=\dot{x} p-L=\frac{\mathrm{i}}{2}
m\dot{x}^2+\varphi=-\frac{\mathrm{i}}{2m}(p+A)^2+\varphi,
\end{equation}
which is proportional to the ``quantum Hamiltonian"
(\ref{Hamiltonian}) if we ``quantize" $p$ as $\hbar\partial_x$.
However, it is not yet obvious how to understand a
Lie-algebra-valued canonical momentum in classical mechanics.

Several aspects of this formalism can be studied further. The
Lagrangian (\ref{non-relativistic}) of our model is simple enough,
so more general cases should be investigated, which is not easy.
Applications of more physical meaning are still lacking. The
possibility of existence of some classical dynamics corresponding
to the generalized quantum system can be carefully examined.
Moreover, natural developments of this formalism, including
quantum field theory, quantum statistical mechanics, quantum
information and so on, are worthy of future work.

\bigskip

This work is partly supported by the National Natural Science
Foundation of China under Grant No. 10347148.


\begin{thebibliography}{99}

\bibitem{Jordan} P. Jordan, Z. Phys. \textbf{80} (1933) 285;
P. Jordan and E. P. Wigner, Ann. Math. \textbf{35} (1934) 29.

\bibitem{Finkelstein1} D. Finkelstein et al., \emph{Notes on Quaternion Quantum Mechanics}
(CERN, Report 59-7), in C. Hoker, ed. Logico-Algebraic-Approach to
Quantum Mechanics II (Reidel, Dordrecht, 1979).

\bibitem{Finkelstein2} D. Finkelstein et al., J. Math. Phys. \textbf{3} (1962) 207;
ibid. \textbf{4} (1963) 788.

\bibitem{LAK} S. De Leo and K. Abdel-Khalek, Prog. Theor. Phys. \textbf{96} (1996)
823.

\bibitem{MT} D. Minic and C.-H. Tze, Phys. Lett. B \textbf{581} (2004)
111.

\bibitem{FH} R.P. Feynman and A.R. Hibbs, \emph{Quantem Mechanics and Path Integral},
McGraw-Hill (New York, 1965).

\bibitem{Wong} S.K. Wong, Nuovo Cim. A \textbf{65} (1970) 689.

\bibitem{BSSW} A.P. Balachandran, Per Salomonson, Bo-Sture Skagerstam, Jan-Olof
Winnberg, Phys. Rev. D \textbf{15} (1977) 2308.

\bibitem{Hatfield} B. Hatfield, \emph{Quantum Field Theory of Point Particles and
Strings} (Addison-Wesley, Reading, 1992).

\end{thebibliography}
\end{document}